\documentclass[12pt]{article}
\usepackage{epsf}
\usepackage{graphicx}
\usepackage{amsfonts}
\usepackage{subfigure}

\def\e{{\rm e}}

\newcommand{\be}{\begin{equation}}
\newcommand{\ee}{\end{equation}}
\newcommand{\bea}{\begin{eqnarray}}
\newcommand{\eea}{\end{eqnarray}}

\newcommand{\complex}{{{\rm I} \kern -.59em {\rm C}}}

\begin{document}
\begin{titlepage}
  \renewcommand{\thefootnote}{\fnsymbol{footnote}}

  \vskip 0pt plus 0.4fill

  \begin{center}
    \textbf{\LARGE Cosmology of the very early universe}
  \end{center}

 \vskip 1ex plus 0.7fill

  \begin{center}
    {\large E. Galindo Dellavalle$^b$,
    G.~Germ{\'a}n$^a$%
    \footnote{
    E-mail: g.german1@physics.ox.ac.uk},
    A. de la Macorra$^c$%
    \footnote{E-mail: macorra@fisica.unam.mx}\\[0.3cm]
    }
    \textit{
  $^{(a)}$The Rudolf Peierls Centre for Theoretical Physics, University of Oxford, 1 Keble Road, Oxford, OX1 3NP, UK\\[0.3cm]
  $^{(b)}$Abacus College, Threeways House, George Street, Oxford, OX1 2BJ, UK\\[0.3cm]
  $^{(c)}$Instituto de F\'{\i}sica, Universidad Nacional Aut{\'o}noma de
  M{\'e}xico,\\
  Apdo. Postal 20-364, 01000 M{\'e}xico D.F. M{\'e}xico}\\

    \vspace{1ex}
    \vskip 1ex plus 0.3fill

    {\large \today}

    \vskip 1ex plus 0.7fill

    \textbf{Abstract}
  \end{center}
  \begin{quotation}
We study cosmological solutions for the very early universe beginning at the Planck scale for a universe containing radiation, curvature and, as a simplification of a possible scalar 
field potential, a cosmological constant term. The solutions are the natural counterpart of the well known 
results for a post-inflationary universe of non-relativistic matter, curvature and a cosmological constant. 
Contrary to the common belief that inflation arises independently of the initial curvature we show that in the 
positive curvature case the universe collapses again into a Big Crunch without allowing the cosmological term to 
dominate and to produce inflation. There is a critical value for the cosmological constant which divide the 
regions where inflation is allowed from those where inflation cannot occur. One can also have 
loitering solutions where the scale factor remains almost constant growing to produce inflation (or decreasing to 
a Big Crunch) after a time which depends on the amount of energy above (or below) the critical energy. At the 
critical energy the solution approaches asymptotically a particular value for the scale factor (Einstein's 
static pre-inflationary universe). The cases where the cosmological term vanishes or becomes negative are 
also studied providing a complete discussion of Friedmann models.
\end{quotation}

\vskip 0pt plus 2fill

\setcounter{footnote}{0}

\end{titlepage}

\section{Introduction}\label{intro}

Studies of the evolution of the universe typically deal with the radiation and matter dominated eras in the old 
cosmology and with inflation and quintessence in the modern standard cosmology scenario. However, the 
pre-inflationary epoch is clearly also of great interest since it may contain the link between more fundamental 
theories (strings, branes) and low energy physics. With this in mind we have made a simplified attempt to 
understand the evolution of the universe in a pre-inflationary era. We study the problem of the evolution of 
the very early universe in the presence of radiation, curvature and a cosmological constant. The cosmological 
term is a very natural simplification of the more realistic and difficult problem of dealing with the evolution 
of a scalar field. The problem we discuss below is based on an exact solution to Friedmann equations. This solution 
is closely related to the one originally obtained by Harrison \cite {harrison}. However to the best 
of our knowledge no further study of such solution has been published perhaps because interest has centred mostly 
in inflationary and post-inflationary solutions. This solution is the natural counterpart of the classical 
post-inflationary problem which has been discussed since long ago \cite {weinberg} and which deals with a universe 
containing non-relativistic matter, curvature and a cosmological term.

\section{A Solution to Friedmann equations} \label{model}

We begin by writing the equations of Friedmann and acceleration in a convenient way
\be
\frac{{\dot R}^2}{R^2} = \frac{8\pi G}{3}\left(\rho_{\gamma}+\rho_{\Lambda}-\rho_k\right) ,
\label{fr}
\ee

\be
\frac{\ddot R}{R} = \frac{8\pi G}{3}\left(-\rho_{\gamma}+\rho_{\Lambda}\right) ,
\label{ac}
\ee
where
\be
\rho_{\gamma} = \rho_{*\gamma}\left(\frac{R_*}{R}\right)^4 , 
\quad \rho_{\Lambda} = \frac{\Lambda}{8\pi G} \equiv \frac{{\varepsilon}^2}{G} ,
\quad \rho_k = \frac{3 k c^2}{8\pi G R^2} ,
\label{ros}
\ee
are the densities of radiation, cosmological constant and curvature, respectively.
The obvious time dependence has been dropped for clarity of notation. As usual $G$ 
denotes Newton's constant and $c$ is the velocity of light. Adding Eq.(\ref{fr}) and 
Eq.(\ref{ac}) and multiplying by $R^2$ we get
\be
R\ddot R + {\dot R}^2 = \frac{16\pi G}{3}\rho_{\Lambda} R^2- k c^2.
\label{frac}
\ee
We notice that the $\it{l.h.s}$ of Eq.(\ref{frac}) can be written as 
\be
R\ddot R + {\dot R}^2 = \frac{1}{2}\frac{d^2}{dt^2} R^2.
\label{identity}
\ee
Introducing the new variable
\be
X = R^2 - 2kc^2t^2_{\Lambda},
\label{ex}
\ee
where
\be
t_{\Lambda} = \sqrt{\frac{3}{32\pi G\rho_{\Lambda}}}.
\label{telam}
\ee
We see that Eq.(\ref{frac}) can be rewritten in the very simple form
\be
\ddot X = t^{-2}_{\Lambda} X.
\label{exeq}
\ee
Thus the solution for $R$ is given by
\be
R = \left(c_1 \e^{-\frac{t}{t_{\Lambda}}} + \frac{1}{2} c_2 \e^{\frac{t}{t_{\Lambda}}} 
+ 2kc^2t^2_{\Lambda}\right)^{1/2}.
\label{sol1}
\ee
The arbitrary constants $c_1$ and $c_2$ are, as usual, fixed by the initial conditions on 
$R$ and on $\dot R$ by using Friedmann equation. Requiring that $R(t = t_*) = R_*$, where the 
asterisk signals some initial condition, gives
\be
c_1 = R^2_* \e^{\frac{t_*}{t_{\Lambda}}}-\frac{1}{2} c_2 \e^{\frac{2 t_*}{t_{\Lambda}}} - 
2t^2_{\Lambda} k c^2 \e^{\frac{t_*}{t_{\Lambda}}}.
\label{c1}
\ee
From the Friedmann equation Eq.(\ref{fr}) evaluated at $t = t_*$ we get the second constant
\be
c_2 = R^2_* \e^{-\frac{t_*}{t_{\Lambda}}}-2t^2_{\Lambda} k c^2 \e^{-\frac{t_*}{t_{\Lambda}}} 
+ \left(\frac{\rho_{*\gamma}}{\rho_{\Lambda}}  
-\frac{4 t^2_{\Lambda} k c^2}{R^2_*}+1 \right)^{1/2} R^2_*\e^{-\frac{t_*}{t_{\Lambda}}}.
\label{c2}
\ee
Thus the solution is given by
\be
R = R_* \left(\cosh \tau + \frac{2t^2_{\Lambda} k c^2}{R^2_*}(1-\cosh \tau) + 
\left(\frac{\rho_{*\gamma}}{\rho_{\Lambda}}  - 
\frac{4 t^2_{\Lambda} k c^2}{R^2_*} + 1 \right)^{1/2} \sinh \tau \right)^{1/2} ,
\label{sol2}
\ee
where we have introduced the new variable $\tau$ defined in terms of $t$ by
\be
\tau \equiv \frac{t-t_*}{t_{\Lambda}} .
\label{tau}
\ee
We can simplify Eq.(\ref{sol2}) by introducing the (constant) density parameters
\be
\Omega_{*i} \equiv \frac{\rho_{*i}}{\rho_{*c}}\equiv \frac{\rho_{i}(t=t_*)}{\rho_{c}(t=t_*)} ,
\label{omegai}
\ee
where $\rho_{*c}$ is the critical density at $t = t_*$ and $\rho_{*i}$ the energy density 
of the substance $i$ at $t = t_*$. We also use the Friedmann equation written in terms of density 
parameters evaluated at $t = t_*$
\be
\Omega_{*\gamma} + \Omega_{*\Lambda} - \Omega_{*k} = 1 .
\label{from}
\ee
We then have the final expression for our solution which is valid for any $k$ and $\Lambda$

\be
R = R_* \left( \frac{1}{2}\frac{\Omega_{*k}}{\Omega_{*\Lambda}} 
+ \left(1 - \frac{1}{2}\frac{\Omega_{*k}}{\Omega_{*\Lambda}}\right)\cosh \tau 
+ \frac{1}{\sqrt{\Omega_{*\Lambda}}}\sinh \tau \right)^{1/2} .
\label{sol}
\ee
Closely related solutions were originally obtained by Harrison \cite {harrison} case by case. 
Here we concentrate in the cosmological consequences of such a solution. The Hubble parameter 
$H \equiv \frac{\dot R}{R}$ is
\be
H = \frac{1}{2 t_{\Lambda}}\left(\frac{R_*}{R}\right)^2 
\left[\left(1 - \frac{1}{2}\frac{\Omega_{*k}}{\Omega_{*\Lambda}}\right)\sinh \tau 
+ \frac{1}{\sqrt{\Omega_{*\Lambda}}}\cosh \tau \right] .
\label{hubble}
\ee
The velocity $\dot R$ is given by
\be
\dot R = H R ,
\label{rp}
\ee
and the acceleration $\ddot R$ is then
\be
\frac{\ddot R}{R} = \frac{1}{2 t^2_{\Lambda}}\left(1-\frac{1}{2}\frac{\Omega_{*k}}{\Omega_{*\Lambda}}
\left(\frac{R_*}{R}\right)^2\right) - \frac{H}{2 t_{\Lambda}}\left(\frac{R_*}{R}\right)^2 .
\label{rpp}
\ee
The case when $\rho_{\Lambda}=0$ follows from an expansion of Eq.(\ref{sol}) on $\tau$ and trowing away 
terms proportional to $\rho_{\Lambda}$ (contained in $t_{\Lambda}$, see Eq.(\ref{telam})). We can also follow 
the previous procedure. In any case the solution is
\be
R = R_*\left(1 + 2 H_*(t-t_*) - \Omega_{*k}H_*^2(t-t*)^2\right)^{1/2} ,
\label{solzero}
\ee
where
\be
H^2_* = \frac{8\pi G}{3}\left(\rho_{*\gamma} - \rho_{*k}\right) .
\label{hubblezero}
\ee
Finally the solution for a negative cosmological constant also follows from Eq.(\ref{sol})
\be
R = R_* \left(-\frac{1}{2}\frac{\Omega_{*k}}{\Omega_{*\Lambda}} 
+ \left(1 + \frac{1}{2}\frac{\Omega_{*k}}{\Omega_{*\Lambda}}\right)\cos \tau 
+ \frac{1}{\sqrt{\Omega_{*\Lambda}}}\sin \tau \right)^{1/2} .
\label{solnegative}
\ee
From now on we particularize to the case where the initial values correspond to the Planck era
\be
\left(t_* , R_* , \rho_{*\gamma}\right) = \left(t_{pl} , l_{pl} , \rho_{pl}\right) ,
\label{plancks}
\ee
where $t_{pl}, l_{pl}$ and $\rho_{pl}$ are Planck time, Planck lenght and Planck density, respectively.
We further work in units where $\hbar = c = m_{pl} = 1$ so that
\be
\left(t_* , R_* , \rho_{*\gamma} , G\right) = \left(t_{pl} , l_{pl} , \rho_{pl} , m^{-2}_{pl}\right) 
= \left(1 , 1 , 1 , 1\right) .
\label{normalizedplancks}
\ee
Thus the densities are given by
\be
\left(\rho_{*\gamma} , \rho_{*\Lambda} , \rho_{*k} , \rho_{*c}\right) 
= \left(1 , \varepsilon^2 , \frac{3k}{8\pi} , 1 + \varepsilon^2 - \frac{3k}{8\pi}\right) .
\label{densitiesplanck}
\ee
We introduce the quantity $b(t)$ which by analogy with the usual $a(t)$ is here defined 
by
\be
b(t) \equiv \frac{R}{R_*} ,
\label{sf}
\ee
where $b_* = b(t=t_*) = b_{pl} = 1$. The notation reminds us that $b(t)$ refers to an 
epoch "before" inflation while the usual $a(t)$ refers to the evolution of the scale factor 
"after" inflation.

\section{Friedmann models}\label{results}

A typical value for the inflationary scale is ${\cal O}(10^{15} GeV)$ thus, to illustrate the 
behaviour of $b(t)$ and all the other quantities of interest we take $\varepsilon = 10^{-8}$. 

$\bullet$ {\bf Positive cosmological term.} In $Fig. 1$ we show the behaviour of $b(t)$ as given 
by Eq.(\ref{sol}) for the three possible values of the curvature.
\begin{figure}[t]
\centering
\centerline{\epsffile{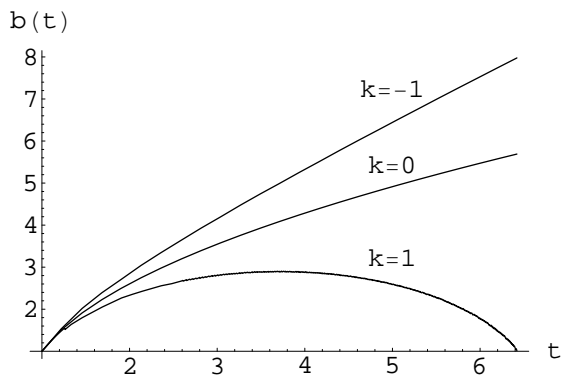}}
\caption{The solution Eq.(\ref{sol}) for a universe containing radiation, curvature and a positive cosmological 
term is here shown as a function of time. All quantities are in Planck units. We see that the positive curvature 
case leads to a Big Crunch shortly after the birth of the universe not allowing the possibility of inflation. 
Inflation can occur in this case only if the energy density of the cosmological term is bigger than a critical 
value, see Eq.(\ref{varepsilonc})}
\label{fig1}
\end{figure}
The case $k = 1$ (closed universe) presents a Big Crunch (BC) at a time very close to the Planck time 
not allowing the possibility of an inflationary epoch. For the open and flat universes $k=-1, k=0$, 
respectively inflation is unavoidable due to the dominance of the $\rho_{\Lambda}$ term at large $b(t)$. 
The case $k=1$ is particularly interesting. We can calculate, as a function of $\Omega_{*\Lambda}$, the 
time at which the BC occurs. This is given by
\be
\tau = \cosh^{-1} \left(1 + \frac{8\Omega_{*\Lambda}}{\left(\Omega_{*k}-2\Omega_{*\Lambda}\right)^2 - 
4\Omega_{*\Lambda}}\right) .
\label{taubc}
\ee
From Eq.(\ref{taubc}) we see that the BC disappears when
\be
\left(\Omega_{*k}-2\Omega_{*\Lambda}\right)^2 - 4\Omega_{*\Lambda} = 0 ,
\label{eqtaubc}
\ee
or
\be
\Omega_{*\Lambda c} = \frac{1}{2}\left(1 + \Omega_{*k} - \sqrt{1 + 2\Omega_{*k}}\right) ,
\label{omegac}
\ee
where $k=+1$ and the subindex $c$ refers to a critical value. We can also solve in terms of $\varepsilon$ 
with the result
\be
\rho_{\Lambda c} = \frac{1}{4}\frac{\rho^2_{* k=1}}{\rho_{*\gamma}} = \left(\frac{3}{16\pi}\right)^2 
= \varepsilon^2_c  .
\label{varepsilonc}
\ee
With this value of $\rho_{\Lambda c} = \varepsilon^2_c$ we find that the solution Eq.(\ref{sol}) reduces to
\be
b_c(t) = \frac{1}{\sqrt{\varepsilon_c}}\left(1+(\varepsilon_c - 1)\e^{-\tau}\right)^{1/2} .
\label{solc}
\ee
\begin{figure}[t]
\centering
\centerline{\epsffile{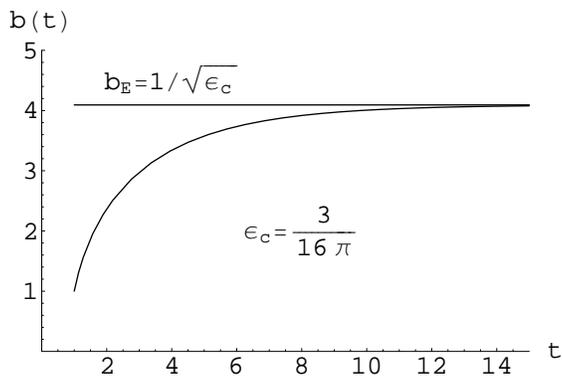}}
\caption{At the critical density Eq.(\ref{varepsilonc}) the solution for a positive cosmological term 
Eq.(\ref{sol}) reduces to Eq.(\ref{solc}). We plot this solution as a function of time $t$. The value it 
approaches $b_E = 1/\sqrt{\varepsilon_c}$ corresponds to a static Einstein solution for a pre-inflationary universe. 
This mini universe is of the order of the Planck lenght.}
\label{fig2}
\end{figure}
This solution is illustrated in $Fig. 2$, where we see that it approches asymptotically the 
value $b_E = 1/\sqrt{\varepsilon_c}$. We can recognize $b_E$ as an Einstein solution for a 
static universe in the pre-inflationary era. The size of this mini universe is of the order of the 
Planck lenght.
\begin{figure}[t]
\centering
\centerline{\epsffile{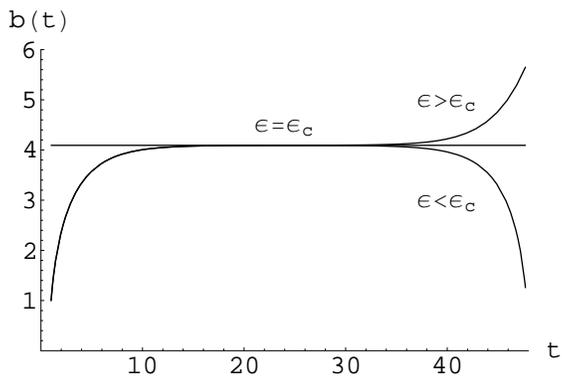}}
\caption{Solutions where the energy density is close to the critical density Eq.(\ref{varepsilonc}) are 
here shown as functions of time. For an energy $\varepsilon < \varepsilon_c$ the solution collapses into a 
Big Crunch while the opposite case gives rise to an inflationary solution with the scale factor growing at an 
ever increasing velocity. By tuning the energy close to the critical value we can have loitering 
solutions where the scale factor remains almost constant during an arbitray interval of time}
\label{fig3}
\end{figure}
As discussed long ago for a post-inflationary universe, a slightly bigger value for the energy density 
than the critical one $\varepsilon_c $ gives rise to a universe which, although closed, 
expands with an ever increasing velocity due to the dominance of the cosmological term. 
On the other hand an energy scale less than $\varepsilon_c$ leads irremediably to a BC. 
This behaviour is illustrated in $Fig. 3$.
We see that we can keep the solution waiting close to $b_E$ as long as we wish (loitering solution) by 
tuning the energy density to values close to the critical value. Loitering solutions in standard cosmology 
for a closed FRW model with matter and a cosmological constant were originally studied by Lema\^itre \cite {lema}. 
Some other more general studies can be found in \cite {otros}.
The behaviour of the Hubble parameter is peculiar in loitering solutions. This has been noted by Sahni 
in the context of loitering braneworld models \cite {sahni}. There, he notices that loitering solutions are characterized 
by the fact that the Hubble parameter dips in value during loitering. We illustrate this behaviour in $Fig. 4$ (long 
dashed curve). The acceleration in a loitering solution always becomes positive for large $b(t)$ when the cosmological 
constant term becomes dominating. The transition from a decelerating to an accelerating universe is signaled by the
point at which the acceleration vanishes. For our solution Eq.(\ref{sol}) (with $k=1$) this is given by
\be
\tau_0 = arc\cosh \left[\frac{9-16 \pi \epsilon \left(3+\epsilon 
\left(3-16 \pi \epsilon+2 \sqrt{\frac{(3-16\pi\epsilon)(3-8\pi(1+\epsilon^2))}{\epsilon}} \right) \right)}
{9-256 \pi^2 \epsilon^2}\right] .
\label{taucero}
\ee 
\begin{figure}[t]
\centering
\centerline{\epsffile{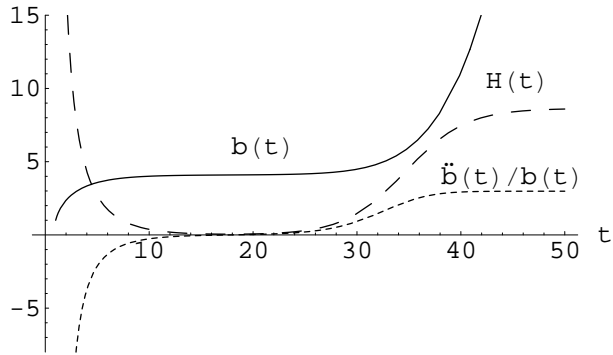}}
\caption{We show the solution Eq.(\ref{sol}) (solid line) for a value of the energy slightly above the critial energy 
$\varepsilon = \varepsilon_c + 10^{-6}$, as well as the Hubble parameter (long dashed line) and the acceleration of 
the scale factor (short dashed). We can see the dip of the Hubble parameter during loitering. To accommodate all the 
curves in one figure we have actually magnified H by a factor of fifty and the normalized acceleration by a factor of 
one hundred.}
\label{fig4}
\end{figure}
\begin{figure}[t]
\centering
\centerline{\epsffile{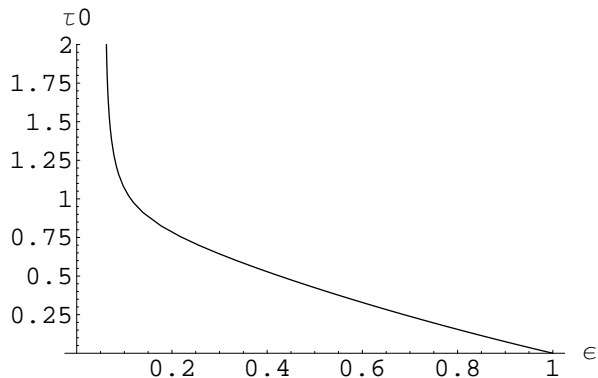}}
\caption{The value of $\tau$ for which the acceleration of the scale factor vanishes is here shown as a function 
of the energy $\varepsilon$ where $\varepsilon_c < \varepsilon < 1$. For $\varepsilon$ = $\varepsilon_c$ the value 
of $\tau_0$ grows without limit corresponding to the asymptotic solution Eq.(\ref{solc}) illustrated in $Fig. 2$. For 
$\varepsilon$ = $1$ (in Planck units), the scale factor accelerates right from the beginning ($\tau_0$ = $0$). The closer 
to $\varepsilon_c$ we tune the energy the longer the loitering epoch. Thus the age of the universe can increase 
dramatically.}
\label{fig5}
\end{figure}
The behaviour of $\tau_0$ as a function of the energy $\varepsilon$ is shown in $Fig. 5$

\begin{figure}[t]
\centering
\centerline{\epsffile{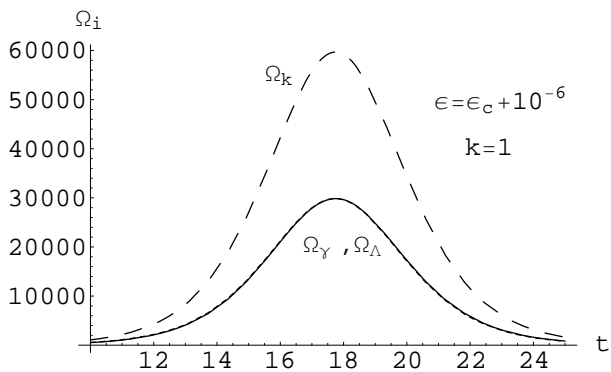}}
\caption{The energy density parameters for the loitering solution with $\varepsilon = \varepsilon_c + 10^{-6}$ and 
$k=1$ are shown as functions of time. We can clearly see that during loitering (compare with $Fig. 4$) the curvature 
term dominates over radiation and the cosmological constant which are almost indistinguishable. For larger values 
of $t$ (not shown) the cosmological term dominates, going asymptotically to one.}
\label{fig6}
\end{figure}
\begin{figure}[t]
\centering
\centerline{\epsffile{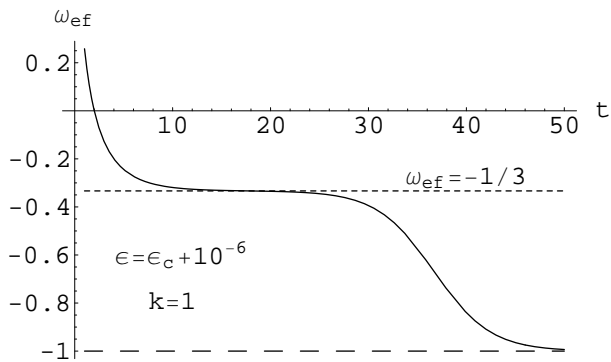}}
\caption{The effective equation of state parameter $\omega_{ef}$ as given by Eq.(\ref{omega}) is shown as a function of time for the loitering 
solution. We see that during loitering $\omega_{ef}\simeq -1/3$ which, consistently with $Fig. 6$, corresponds to 
curvature domination. For larger values $\omega_{ef}$ goes to $-1$ which is the value corresponding to a cosmological 
constant term as it should be.}
\label{fig7}
\end{figure}
\begin{figure}[t]
\centering
\centerline{\epsffile{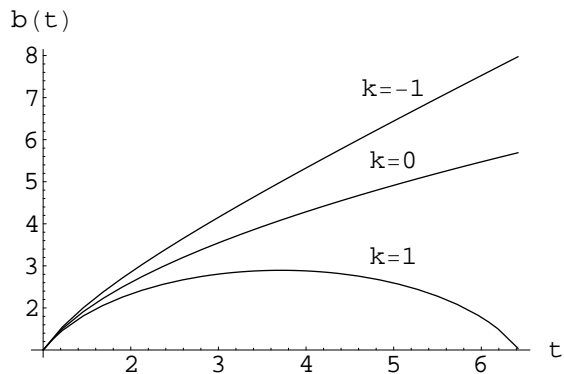}}
\caption{The solution Eq.(\ref{solzero}) for a universe containing radiation, curvature and a vanishing cosmological 
term is here shown as a function of time. The figure is almost identical to $Fig. 1$ because the cosmological 
term is only relevant at large $b(t)$, see $Fig. 9$.}
\label{fig8}
\end{figure}
\begin{figure}[htp!]
\begin{center}
\includegraphics[width=7cm]{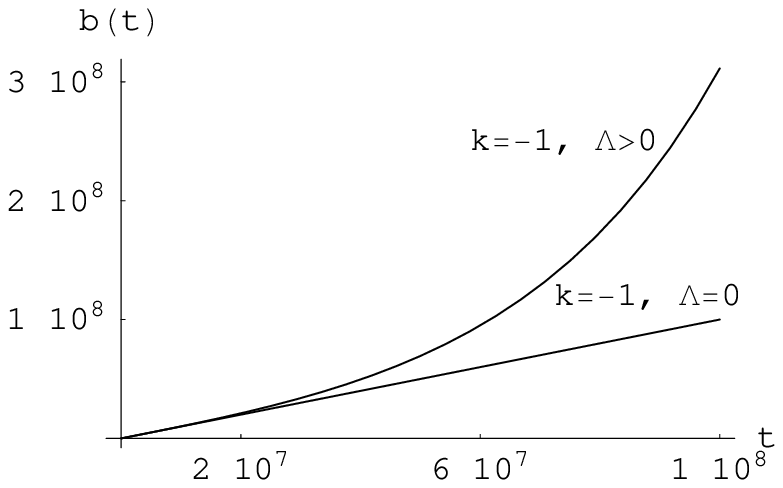}
\includegraphics[width=7cm]{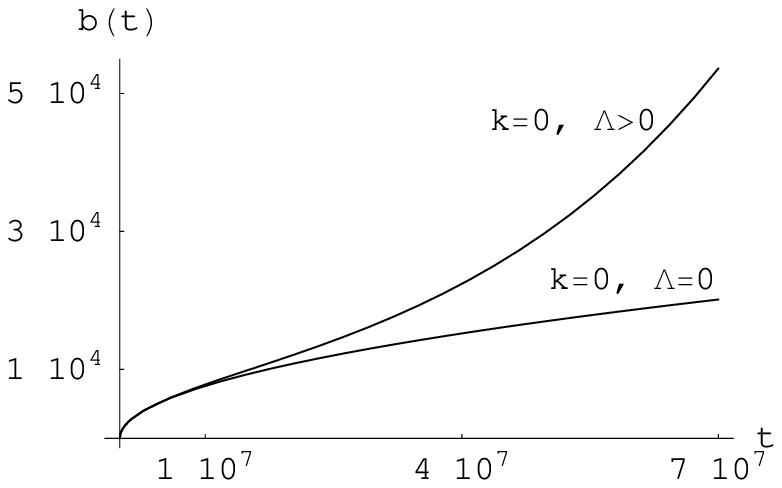}
\end{center}
\caption{Here we show the solution Eq.(\ref{solzero}) for the vanishing cosmological case when $k=-1$ 
and $k=0$. We compare with $Fig. 1$ for large $b(t)$ where the cosmological term begins to dominate 
the expansion of the universe. The case $k=+1$ for $\varepsilon < \varepsilon_c$ is almost identical 
in both cases.}
\label{fig9}
\end{figure} 
We have also investigated the behaviour of the energy density parameters for the loitering solution. This is 
illustrated in $Fig. 6$ where we see that it is the curvature term the one which dominates the energy of the 
universe during loitering, while the cosmological constant and radiation terms remain almost indistinguishable one 
from the other. This peculiar behaviour during loitering is also consistently shown in $Fig. 7$ where we plot the 
effective equation of state parameter $\omega_{ef}$ which takes a value around $-1/3$ during loitering. 
The $\omega_{ef}$ is given by
\be
\omega_{ef}=\frac{\frac{1}{3}\rho_{\gamma}-\rho_{\Lambda}-\frac{1}{3}\rho_{k=1}}{\rho_{\gamma}+\rho_{\Lambda}+\rho_{k=1}}
\label{omega}
\ee

$\bullet$ {\bf Vanishing cosmological term.} In this case the solution is given by Eq.(\ref{solzero}) for 
the three values of $k$. This solution is illustrated in $Fig. 8$. 
Note that this figure is almost identical to $Fig. 1$. The reason is that for small $t$ the cosmological term 
present in Eq.(\ref{sol}) is irrelevant.
Thus solutions Eq.(\ref{sol}) and Eq.(\ref{solzero}) become different for larger values of $t$. This is shown 
in $Fig. 9$ for the cases $k=-1$ and $k=0$ only since the $k=+1$ case is essentially the same in both solutions 
whenever $\varepsilon < \varepsilon_c$.
\begin{figure}[htp!]
\begin{center}
\includegraphics[width=7cm]{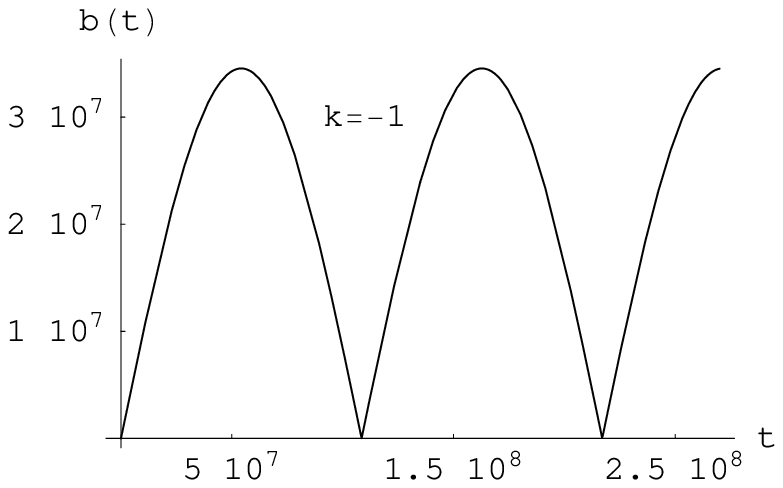}
\includegraphics[width=7cm]{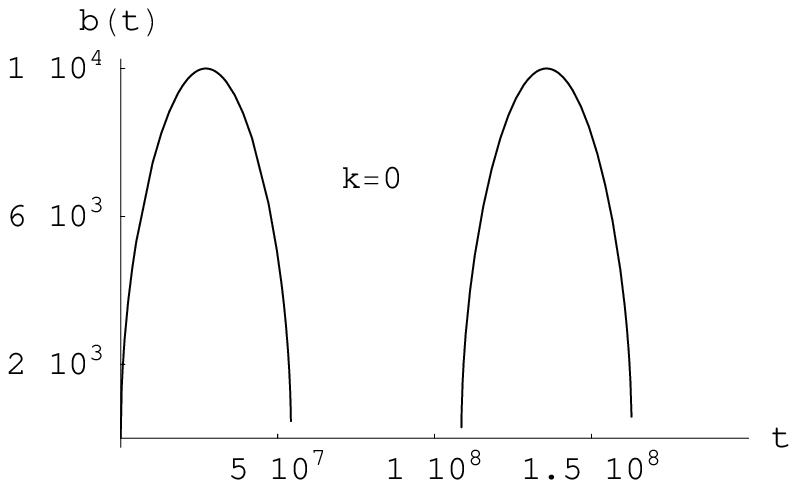}
\includegraphics[width=7cm]{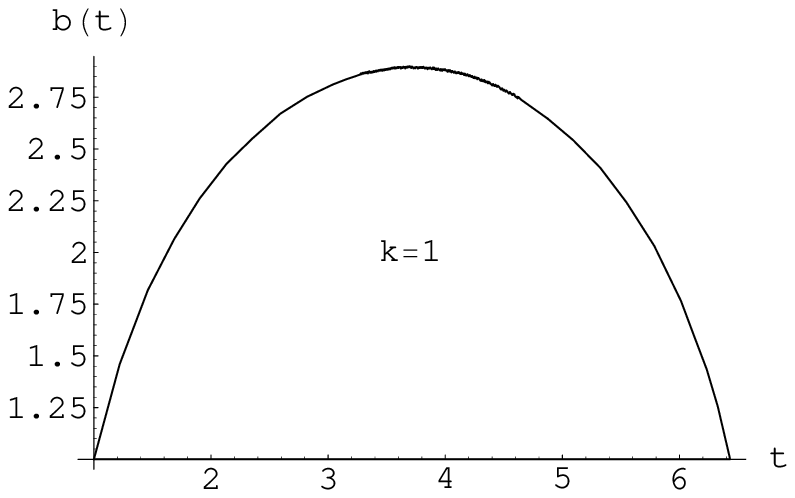}
\end{center}
\caption{We show the solution Eq.(\ref{solnegative}) for a universe containing radiation, curvature and a negative 
cosmological term as a function of time. In all cases the universe collapses into a Big Crunch oscillating in 
different ways for the negative and vanishing curvature cases while collapsing once for the positive curvature. }
\label{fig10}
\end{figure}

$\bullet$ {\bf Negative cosmological term.} In this case the solution is given by Eq.(\ref{solnegative}) for 
the three values of $k$. This solution is illustrated in $Fig. 10$.

\section{Conclusions}\label{con}

We have studied a model of the very early universe containing radiation, curvature and a cosmological constant. 
We find that in the cases $k=-1$, $k=0$ and $\Lambda > 0$ inflation occurs unavoidably while in the case $k=+1$ there 
is no inflation unless the energy density is bigger than a certain critical value. This critical energy 
signals the limit between contracting solutions which end in a Big Crunch from those which expand forever. 
At the critical value, the solution approaches asymptotically an Einstein solution for a static universe 
in the pre-inflationary regime. The mini universe defined by this solution is of the order of the Planck 
lenght. Particular emphasis is placed in the closed universe case because there loitering solutions can occur. 
These solutions have somewhat peculiar properties: the Hubble parameter dips in value during loitering and 
the energy density parameter associated with curvature dominates over radiation and the cosmological constant. 
We also solve and briefly discuss the cases with a vanishing and negative cosmological constant thus 
providing a complete study of the Friedmann models. All of these results have been also obtained in the well 
studied problem of a universe of matter, curvature and a cosmological constant. They are due to the competition 
between curvature and the matter terms. Thus, technically, replacing matter by radiation which decays 
even faster should not change any of these features. However it is most gratifying that in this case one can find 
closed form analytical solutions to the problem.

\section{Acknowledgements}

This work was supported by the project PAPIIT IN114903-3 and CONACYT: 42096, 45718. G.G. is grateful to Prof. G.G. Ross and Prof. S. Sarkar for the hospitality extended to him at the RPCTP during a sabbatical leave from Instituto de Ciencias F\'{\i}sicas, Universidad Nacional Aut\'onoma de M\'exico. Support from DGAPA, UNAM and the RPCTP, Oxford is gratefully acknowledged. This work is part of the Instituto Avanzado de Cosmolog\'{\i}a (IAC) collaboration.

\end{document}